\documentclass{sig-alternate-05-2015}
\usepackage{epstopdf}

\begin{document}

\setcopyright{acmcopyright}

\doi{10.475/123_4}

\isbn{123-4567-24-567/08/06}



%

\title{When Do Luxury Cars Hit the Road? Findings by A Big Data Approach}
%
%
%
%
%

\numberofauthors{2} 
%
\author{
%
%
\alignauthor
Yang Feng\\
       \affaddr{Computer Science}\\
       \affaddr{University of Rochester}\\
       \affaddr{Rochester, NY, 14627}\\
       \email{yfeng23@cs.rochester.edu}
\alignauthor
Jiebo Luo\\
       \affaddr{Computer Science}\\
       \affaddr{University of Rochester}\\
       \affaddr{Rochester, NY, 14627}\\
       \email{jluo@cs.rochester.edu}
}

\maketitle
\begin{abstract}
In this paper, we focus on studying the appearing time of different kinds of cars on the road. This information will enable us to infer the life style of the car owners. The results can further be used to guide marketing towards car owners. Conventionally, this kind of study is carried out by sending out questionnaires, which is limited in scale and diversity. To solve this problem, we propose a fully automatic method to carry out this study. Our study is based on publicly available surveillance camera data. To make the results reliable, we only use the high resolution cameras (i.e. resolution greater than $1280 \times 720$). Images from the public cameras are downloaded every minute. After obtaining 50,000 images, we apply faster R-CNN (region-based convoluntional neural network) to detect the cars in the downloaded images and a fine-tuned VGG16 model is used to recognize the car makes. Based on the recognition results, we present a data-driven analysis on the relationship between car makes and their appearing times, with implications on lifestyles. 
\end{abstract}

%
%


%
%

%
%



\section{Introduction}

\begin{figure}[t]
\centering
\includegraphics[width=3in]{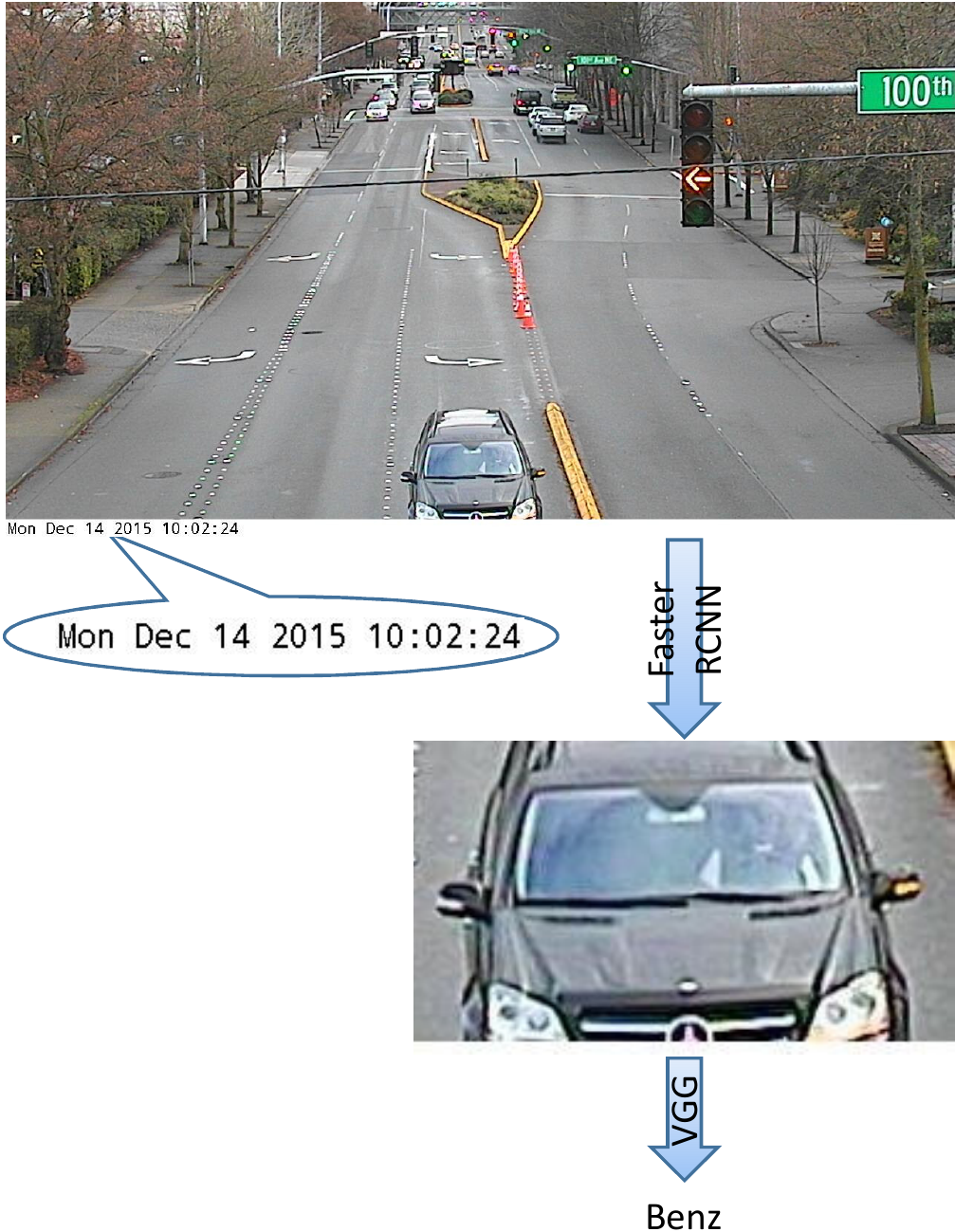}
\caption{The framework of our method. First, we download images from public surveillance cameras at intersections. Second, we detect the cars using Faster R-CNN \cite{ren2015faster}. Third, we recognize the make of the cars using a fine-tuned VGG model \cite{simonyan2014very}. }
\label{fig1}
\end{figure}

It is well known that car ownership rate is very high in America. There were about 1.8 vehicles per U.S. household in 2013 \cite{ownper}. According to the report, the average number of minutes an American spends behind the wheel is 87 per day \cite{usage} and 85\% workers commute to work in a car alone every day. Therefore cars are heavily used in America. From the use of the cars, we can find useful information about the life style of American people. For example, such information can be providence to urban planning or car marketing. 

Traffic jams are common in big cities. One interesting phenomenon about traffic jams is that we rarely see luxury cars in traffic jams. Do luxury car owners know when and where the traffic jam happens and how to avoid traffic jams? To answer this question, Zhang \cite{study} carried out a study in one Chinese community. They manually recorded the leaving time and returning time of the cars in that community everyday. They also recorded the make of each passing car. They found out that the owners of many inexpensive cars usually leave their homes for work early, i.e. at 7 o'clock or 8 o'clock. While the owners of many expensive cars leave their homes no earlier than 9 o'clock. The luxury car owners, however, leave their homes at noon or in the evening. In fact, the luxury car owners usually do not go out during heavy traffic rush hours in the morning and afternoon.

Zhang's study is interesting, but it only reports the result in one community. To draw a more convincing conclusion, their study needs to be carried out in more places. Their method is based on manually counting, which is labor expensive to scale up to many places. Motivated by Zhang's study, we chose to design a fully automatic method to address the same question at scale. 

We use the publicly available surveillance traffic cameras and computer vision methods to facilitate the study. To monitor the traffic condition on the road, the government has installed many surveillance traffic cameras. Some of them are configured as publicly accessible on the Internet. We search for them on Google and find that most of them are low resolution cameras. The low resolution cameras can be used to count how many cars are on the road, but the make of the car cannot be recognized from the images captured by low resolution cameras. Fortunately, upon further search, we find that the Bellevue county of Seattle provides high resolution traffic images captured by surveillance cameras. 
The image for one camera is updated every minute. We choose the high resolution ones and download them.
After obtaining the images, we need to detect the cars and recognize their makes. We use the state-of-the-art object detection method Faster R-CNN \cite{ren2015faster} for detecting cars. To recognize the car make, we fine-tuned the VGG model \cite{simonyan2014very} using the CompCars dataset \cite{yang2015large}. Based on the recognition result, we find that different cars do appear at different times of a day according to their makes. Fig. \ref{fig1} shows the entire framework of our method.

\section{Related Work}

\subsection{Car User Study}
Any work solving the same problem has not been found yet. The most related work we know is Choo and Mokhtarian's research \cite{choo2004type}. Based on a 1998 mail-out/mail-back survey, Choo and Mokhtarian studied the relation between vehicle type choices and lifestyle. The choice of individual's vehicle type is affected by factors including the distance and frequency of trips, how much one enjoy traveling, pro-environmental considerations, family, money, gender and age. Zhang et al. \cite{zhang2015sensing} proposed a data-driven system to analyze individual refueling behavior and citywide petrol consumption. Their method is more efficient and covers more people than the questionnaire-based methods.

\subsection{Car Detection and Recognition}
Ramnath et al. \cite{ramnath2014car} proposed to recognize car make and model by matching 3D car model curves with 2D image curves. During recognition, the car pose is estimated and used to initialize 3D curve matching. Their method is able to recognize the make of a car over a wide range of viewpoints. Fraz et al. \cite{fraz2014mid} represented the vehicle images using a mid-level feature for vehicle make and model recognition. Their mid-level representation is based on SURF and Fisher Vector while an Euclidean distance based similarity is used for classification. These two methods use high resolution car images, while the cars in our images are much smaller. So these methods are not suitable for our problem. 
Recently, He et al. \cite{he2015recognition} have used a part-based detection model to detect cars and designed an ensemble classifier of neural networks to recognize car models. Their method works on a single traffic-camera image and has obtained promising results on their dataset.

\section{Methodology}

\subsection{Car Detection}

We use Faster R-CNN \cite{ren2015faster} to detect cars. This detection method is developed from two previous work, namely R-CNN \cite{girshick2014rich} and Fast R-CNN \cite{girshick2015fast}. The object detection process in R-CNN can be summarized as:

\begin{itemize}
\item Generate object proposals using selective search \cite{uijlings2013selective}.
\item Extract a 4096-dimensional feature vector for each region proposal using AlexNet \cite{krizhevsky2012imagenet}.
\item Score each feature vector using object category classifiers.
\end{itemize}

The AlexNet used in the second step is pre-trained on ILSVRC2012 and fine-tuned on a specific detection data such as VOC or ILSVRC2013. The object category classifiers are trained using SVM. The detection results of R-CNN is good on the VOC dataset and ILSVRC2013 detection test set. However, it is a three-step procedure, in which the features need to be saved and training the object classifiers takes significant time. To solve these drawbacks of R-CNN, Fast R-CNN was proposed. In Fast R-CNN, feeding an image and multiple proposals to a CNN model, the model will output the softmax probabilities and refined bounding box offsets. The detection time reduced significantly in Fast R-CNN, which makes the proposal generating become the bottle neck in object detection. Ren et al. \cite{ren2015faster} introduced Region Proposal Network (RPN) to Faster R-CNN to solve the inefficient proposal generating problem. For an input image, RPN outputs 9 bounding boxes at each pixel location and the scores of how likely a bounding box contains an object. The convolutional layers in RPN is shared with Fast R-CNN, so the convolutional layers only need to be computed one time for proposal generation and object detection. In practice, we use the VGG net released by Ren\footnote{https://github.com/ShaoqingRen/faster\_rcnn}, which was trained on VOC 07/12.

\subsection{Car Recognition}

Convolutional Neural Network (CNN) \cite{lecun1998gradient} is used for car recognition. CNN has obtained good results in many computer vision problems such as image recognition \cite{krizhevsky2012imagenet}, object detection \cite{girshick2014rich}, video classification \cite{karpathy2014large} and face verification \cite{taigman2014deepface}. The success of CNN is due to its ability to learn rich mid-level image features. However, a large amount of data is needed to train a CNN. When there is no sufficient data available for training CNN, some researchers have tried to transfer the image representations learned on image classification task to other tasks and reported promising results \cite{oquab2014learning}. Our problem falls in this kind. We use the pre-trained model for image classification and change the last fully connected layer to output a 163-dimension vector corresponding to the 163 car makes. We fine-tune the VGG model \cite{simonyan2014very} on CompCars dataset \cite{yang2015large}. To keep the image features learned in the layers before the last layer, we use a much smaller learning rate for them.

\section{Data}

\begin{figure}
\centering
\includegraphics[width=2in]{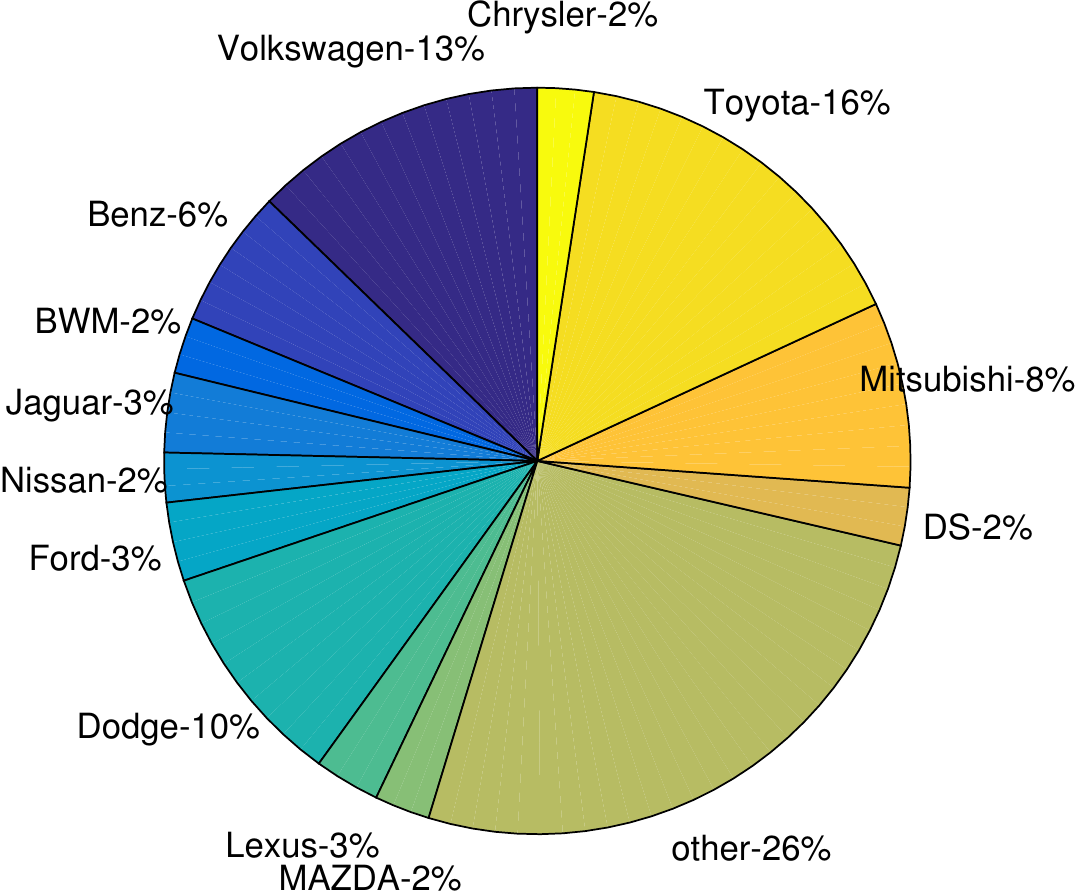}
\caption{\label{fig:to} The makes of all detected cars. The car makes with very small percentages are merged into ``other''.}
\end{figure}

\subsection{The CompCars dataset} The CompCars dataset \cite{yang2015large} was collected by The Chinese University of Hong Kong. This dataset contains 208,826 images of 1,716 car models. These images are divided into web-nature images and surveillance-nature images. All the car models come from 163 car makes. This dataset was originally used for fine-grained car classification, car attribute prediction and car verification.

\subsection{Traffic camera images}
We download the traffic camera images from the website of Bellevue government\footnote{http://www.bellevuewa.gov/trafficmap}. There is a list of cameras available on the website. We choose the ones with high resolution and download the images captured by them. The IDs of the cameras we are using are 003, 042 and 133. The image download URL is "http://www.bellevuewa.gov/TrafficCamImages/\\CCTV\$\{ID\}.jpg", where \$\{ID\} should be replaced by the ID number.

\section{Results}

We exploit the web nature in the dataset to train the car recognition model. There are $136,726$ cars web images in total. We randomly split them into a training part containing $122,995$ images and a testing part containing $13,731$ images. We try the GoogLeNet model \cite{szegedy2014going} and 16 layer VGG model \cite{simonyan2014very}. Table \ref{tab:ra} shows the recognition accuracy on the testing set. VGG model achieves better results, so we use it to recognize the car in the following experiment.

\begin{table}[h]
\centering
\begin{tabular}{|c|c|c|}
\hline
Model & GoogLeNet & VGG \\\hline
Top-1 & 0.844 & 0.933 \\\hline
Top-5 & 0.959 & 0.983 \\\hline
\end{tabular}
\caption{\label{tab:ra}Recognition accuracy on the testing set.}
\end{table}

\begin{figure}
\centering
\includegraphics[width=2in]{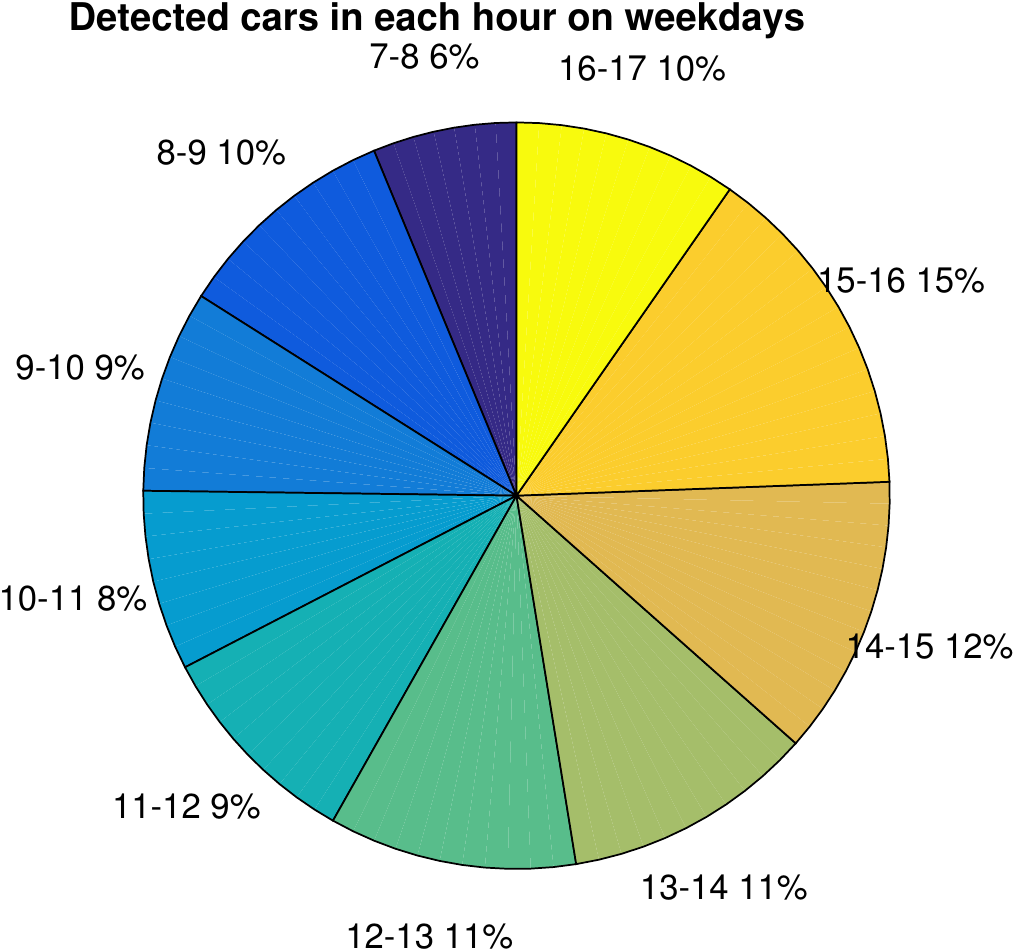}
\includegraphics[width=2in]{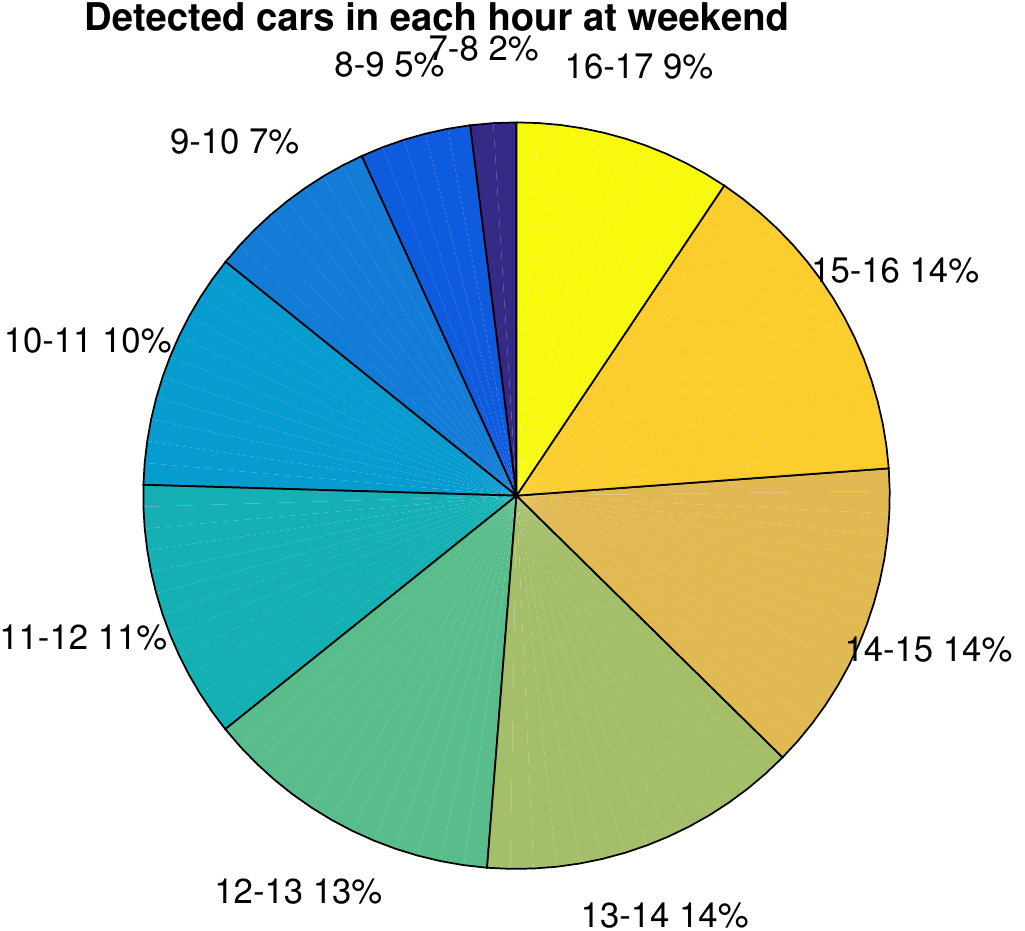}
\caption{The percentages of detected cars in each hour: (top) weekdays and (bottom) weekends.}
\label{fig:ea}
\end{figure}

We remove the images captured from 5 o'clock p.m. to 6 o'clock a.m. because many of them are too dark to recognize. After that, we obtain 49,768 images captured between Nov. 11 and Dec. 18 in 2015. 
Fig. \ref{fig:to} shows the percentage of the car makes of all the detected cars. From the chart, we can find that Toyota, Volkswagen and Dodge are the most popular car makes of the detected cars. Fig. \ref{fig:ea} shows the percent of detected cars in each hour. We observe that nearly half of the cars appear between 7 o'clock and 13 o'clock on weekdays. On weekends, fewer cars go out early in the morning.

\begin{figure}[!h]
\centering
\noindent\makebox[3in][c] {
\includegraphics[width=1.5in]{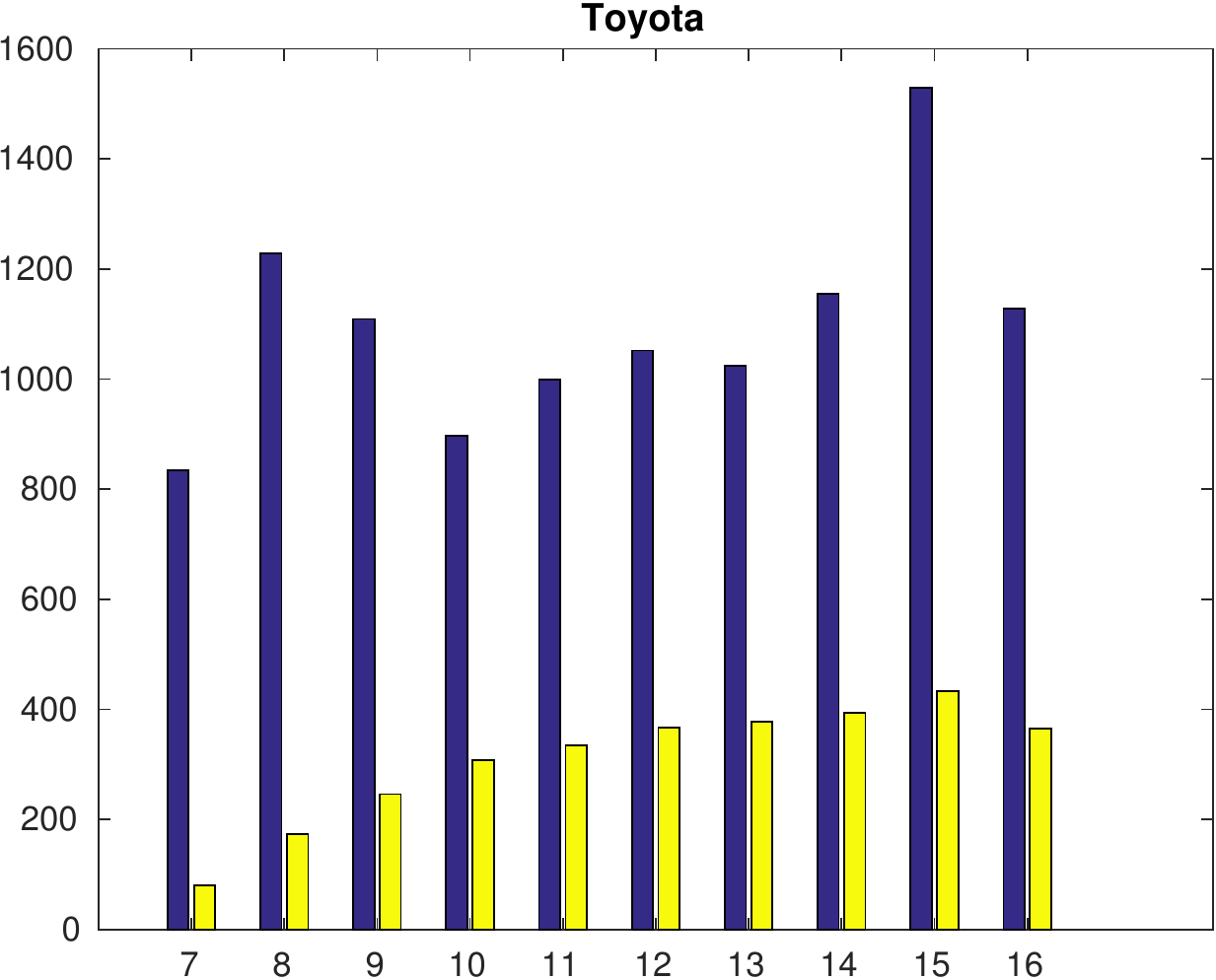}
\includegraphics[width=1.5in]{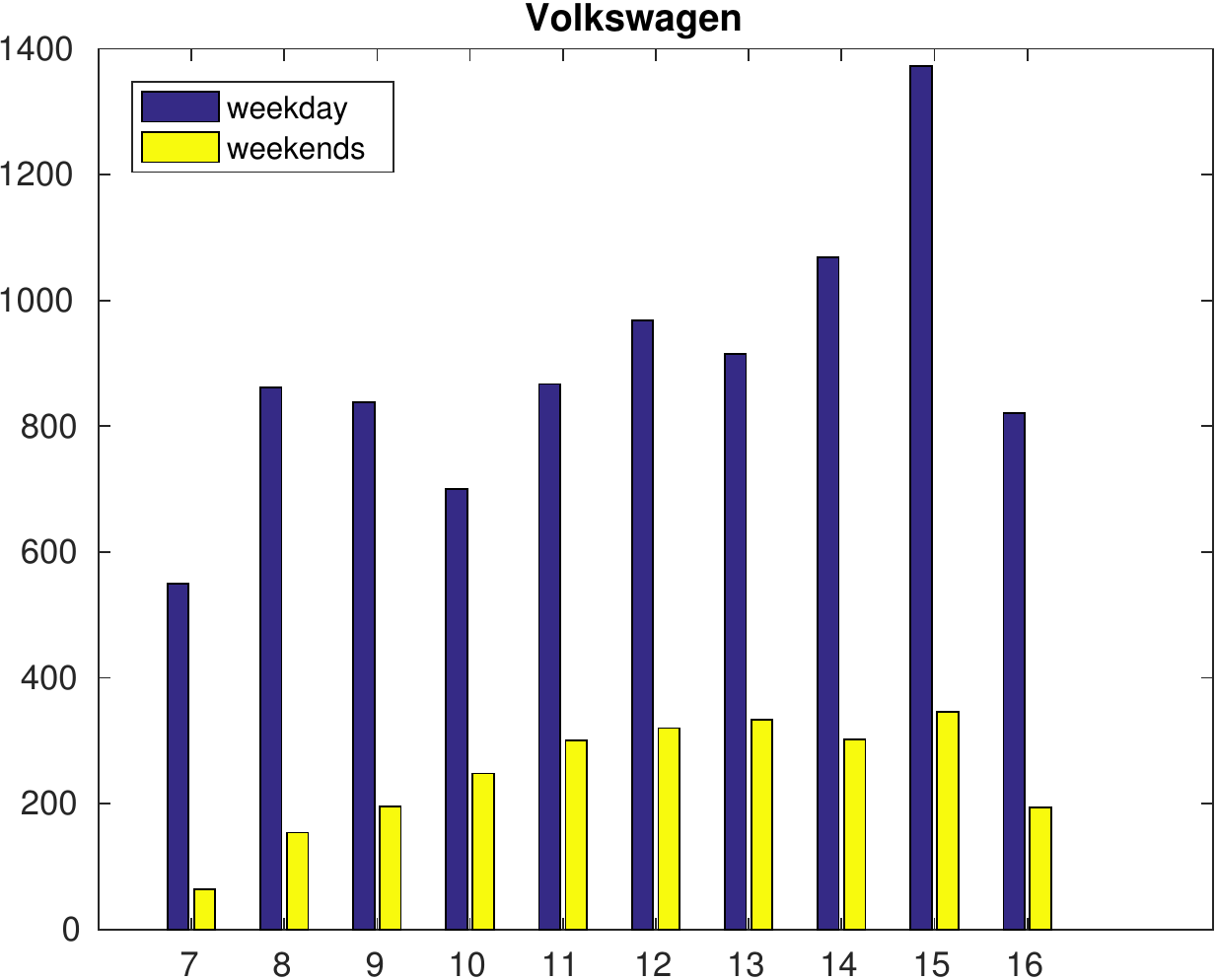}
}
\noindent\makebox[3in][c] {
\includegraphics[width=1.5in]{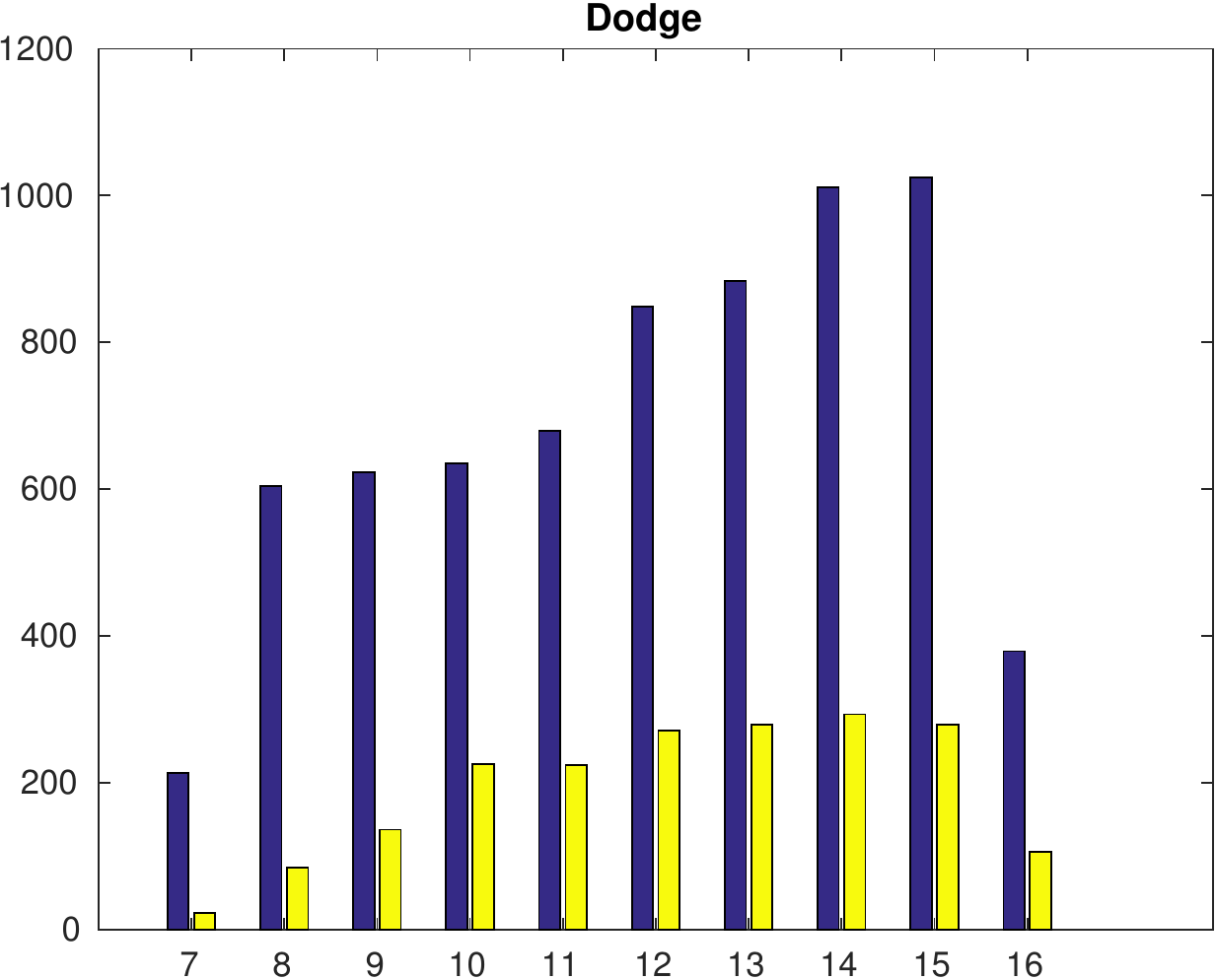}
\includegraphics[width=1.5in]{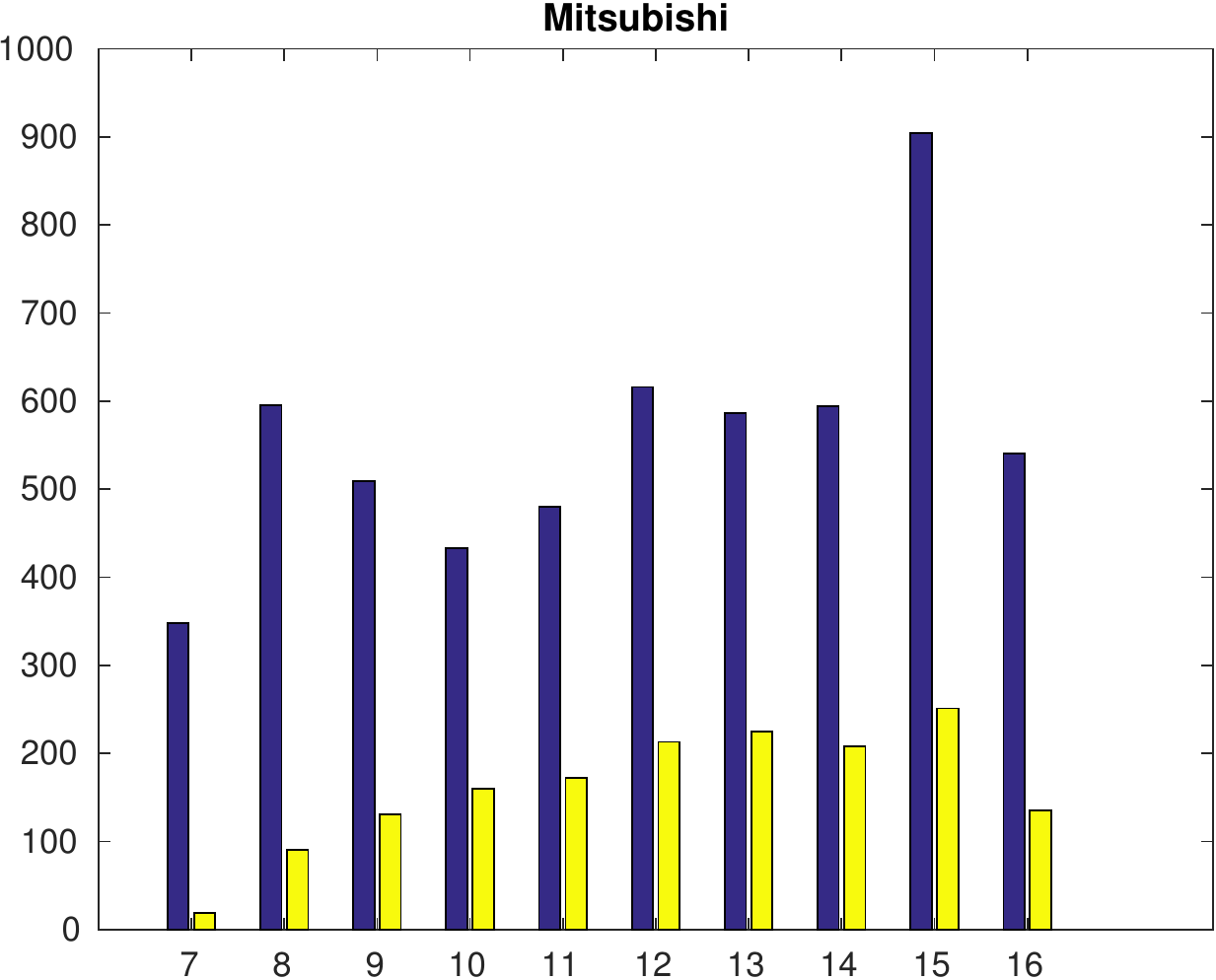}
}
\caption{\label{fig:make1} The appearing hours of Toyota, Volkswagen, Dodge and Mitsubishi.}
\end{figure}

\begin{figure}[!h]
\centering
\noindent\makebox[3in][c] {
\includegraphics[width=1.5in]{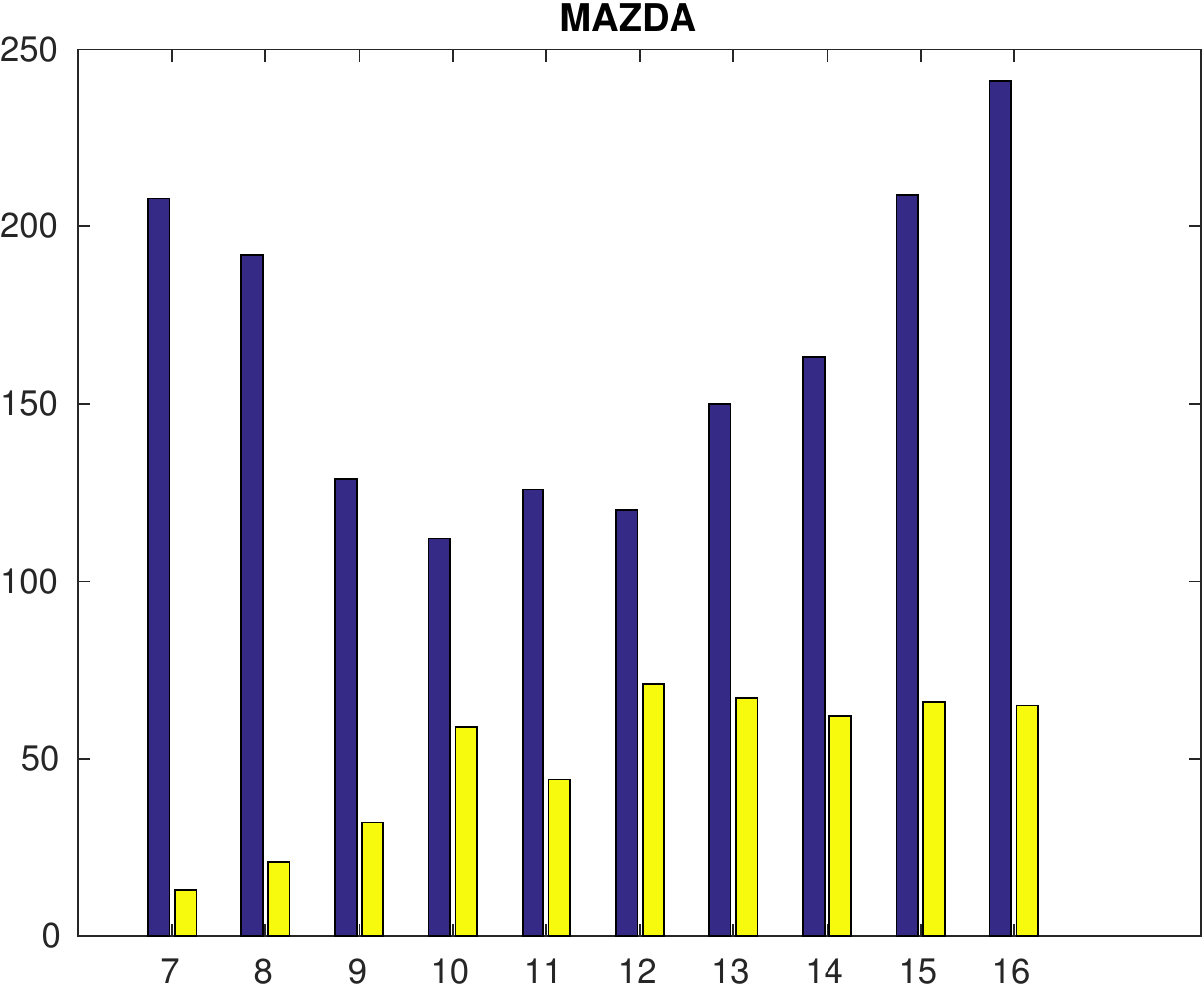}
\includegraphics[width=1.5in]{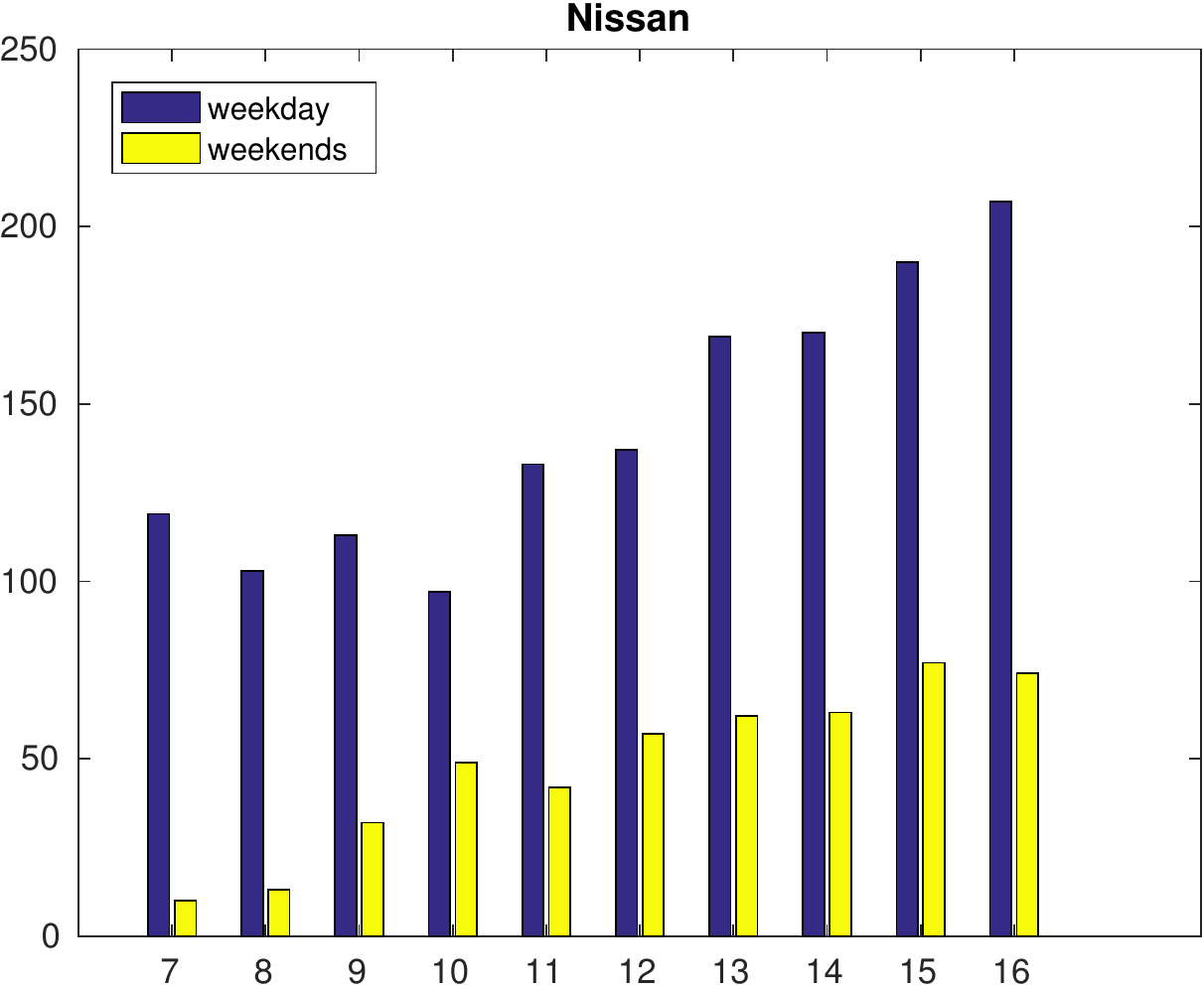}
}
\caption{\label{fig:make2} The appearing hours of MAZDA and Nissan.}
\end{figure}

\begin{figure}[!h]
\centering
\noindent\makebox[3in][c] {
\includegraphics[width=1.5in]{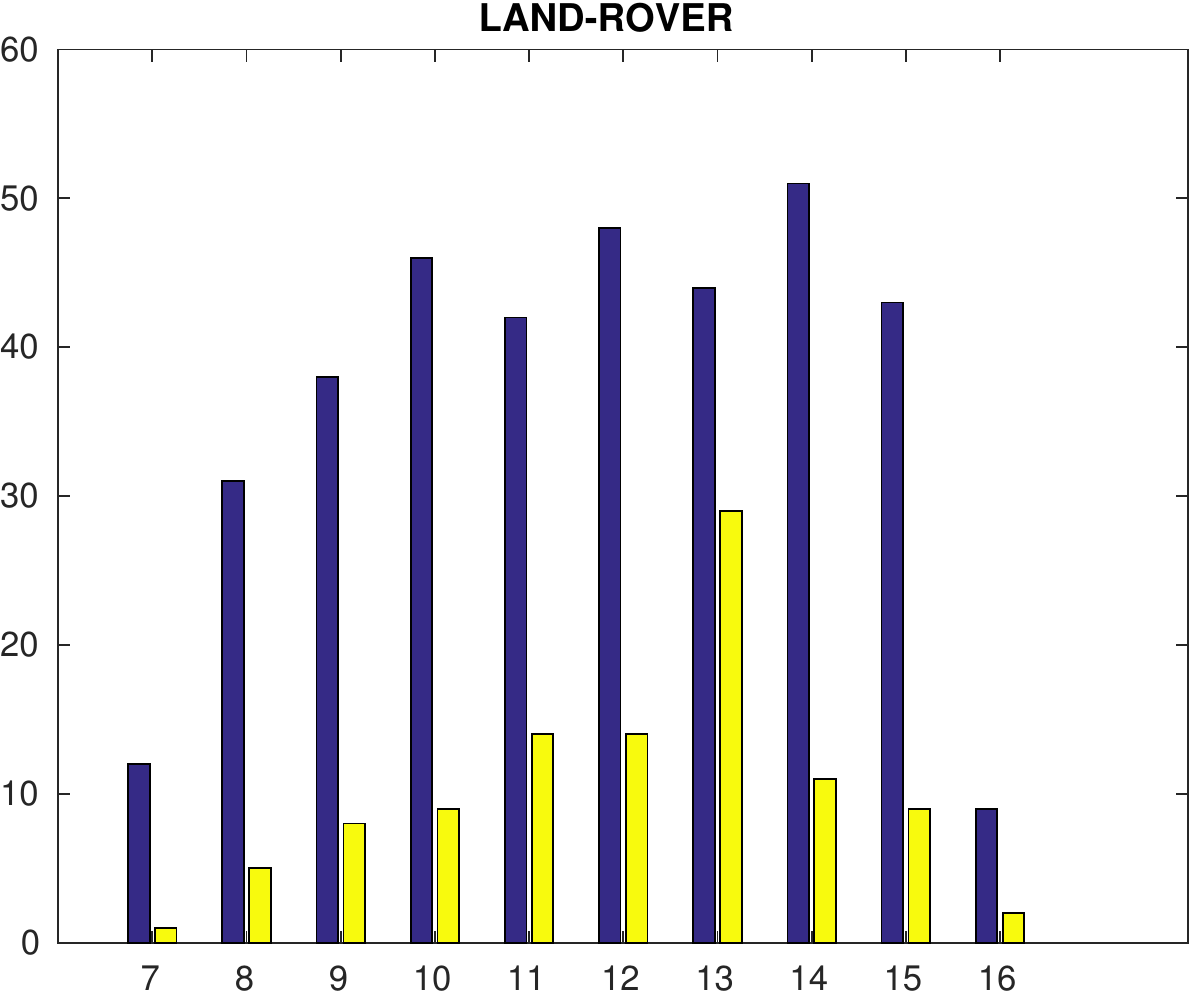}
\includegraphics[width=1.5in]{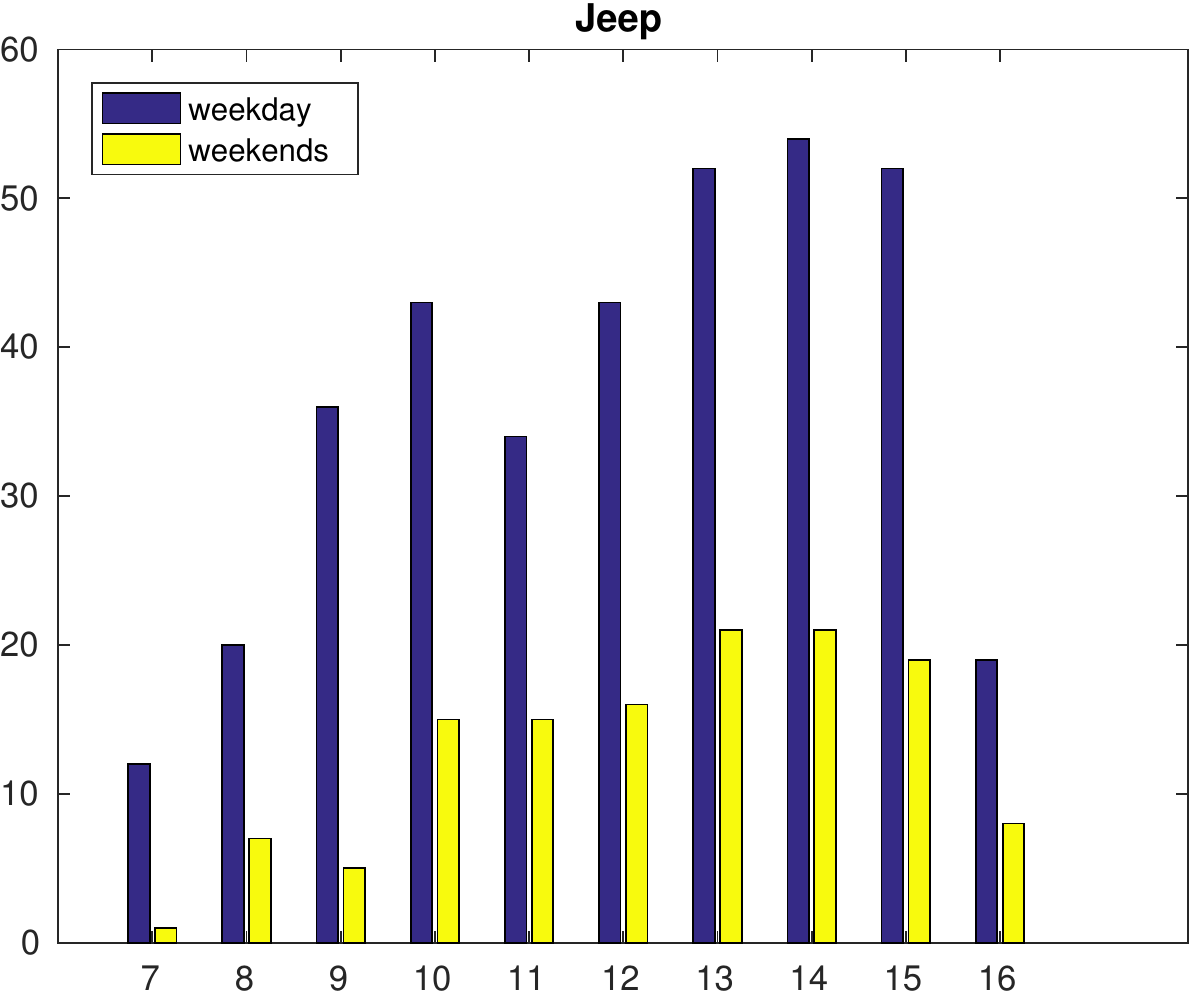}
}
\noindent\makebox[3in][c] {
\includegraphics[width=1.5in]{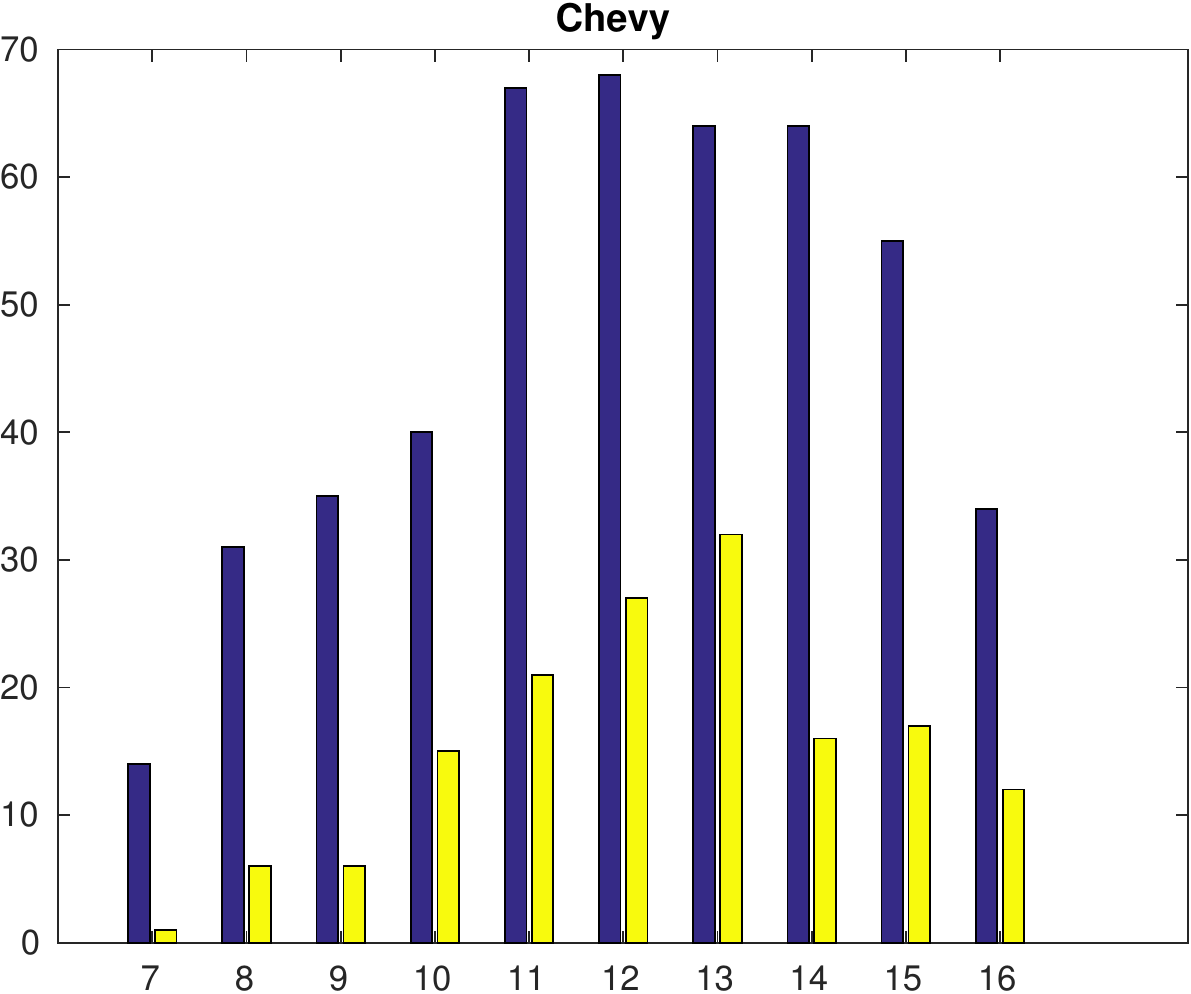}
\includegraphics[width=1.5in]{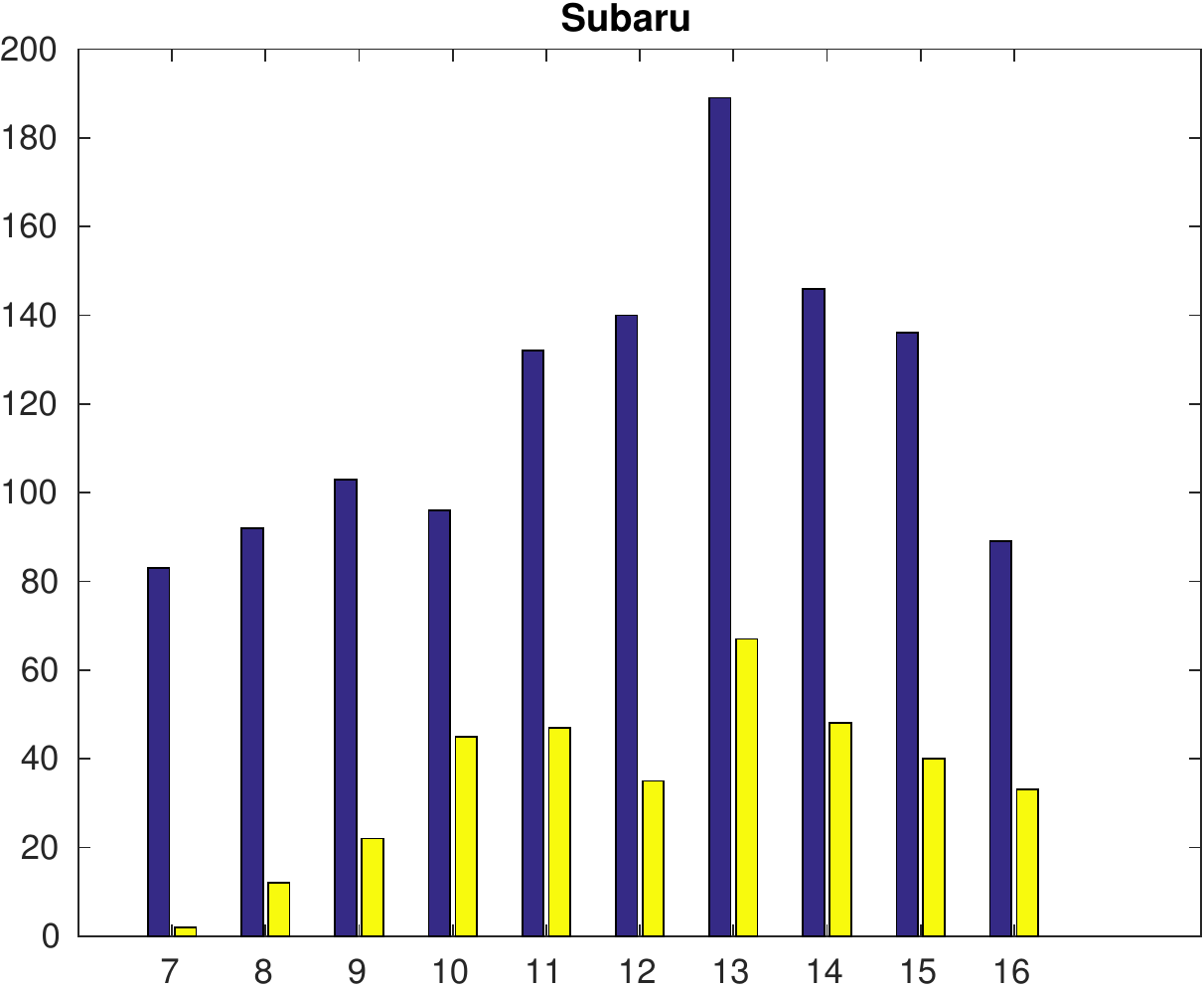}
}
\caption{\label{fig:make3} The appearing hours of LAND-ROVER, Jeep, Chevy and Subaru.}
\end{figure}

The next three figures show the sum of detected cars of some car makes with respect to road hours. Fig. \ref{fig:make1} lists the car makes which are detected most of the time. From Fig. \ref{fig:make1}, we observe that the owners of these four kinds of common cars are likely to go to work at 8 o'clock. Different from them, we can find in Fig. \ref{fig:make2} that more owners of MAZDA and Nissan cars go to work earlier, i.e. at 7 o'clock. The owners of the four types of cars in Fig. \ref{fig:make3} are not early birds. Their driving time does not peak at 7 or 8 o'clock. Moreover, they appear more in the afternoon on weekends, which is also later than the other car owners on weekends.


\section{Conclusion}

We have studied when different classes of cars hit the road using a scalable data-driven approach. We have found that different brands of cars appear in the traffic at different times on weekdays, which suggests that the owners of different cars follow different lifestyles and work schedules. Our future work includes investigating such patterns in different parts of a major city, as well as in different geographic regions with different population demographics. 


\section{Acknowledgment}
This work was supported in part by New York State through the Goergen Institute for Data Science at the University of Rochester, and NSFC grant 61428202.

{
\small
\bibliographystyle{abbrv}
\bibliography{dm.bib}
}

\end{document}